\documentclass[aps,pre,twocolumn,showpacs,preprintnumbers,amsmath,amssymb]{revtex4}

\usepackage{graphicx}

\usepackage{dcolumn}

\usepackage{bm}

\begin{document}

\title{Critical features in electromagnetic anomalies detected prior to the L'Aquila earthquake}

\author{Y.F. Contoyiannis*}

\author{C. Nomicos**}

\author{J. Kopanas+}
\author{G. Antonopoulos+}
\author{L. Contoyianni+}

\author{K. Eftaxias+}
\email{ceftax@phys.uoa.gr}

\affiliation{*Department of Physics and Chemistry, Techological Educational Institute  (TEI) of Chalkis, Chalkis, Greece, **Department of Electronics, Technological Educational Institute of Athens, Ag. Spyridonos, Egaleo, 12210, Athens, Creece, +Department of Physics, University of Athens, GR-15771 Athens, Greece}

\date{\today}

\begin{abstract}

Earthquakes (EQs) are large-scale fracture phenomena in the Earth's heterogeneous crust.
 Fracture induced physical fields allow a real-time monitoring of damage evolution in materials during mechanical loading. 
Electromagnetic (EM) emissions in a wide frequency spectrum ranging from kHz to MHz are produced by opening cracks, 
which can be considered as the so-called precursors of  general fracture. We emphasize that  the MHz radiation appears earlier than  the kHz in both laboratory and geophysical scale. An important challenge in this field of research is to distinguish  characteristic epochs in the evolution of precursory EM activity and identify them with the equivalent  last stages in the EQ preparation process. Recently, we proposed the following two epochs/stages model:  (i) The second epoch, which includes the  finally emerged  strong impulsive kHz EM emission is due to the fracture of the high strength large asperities  that are distributed along the activated fault sustaining the system. (ii) The first epoch,  which includes the initially emerged  MHz EM radiation is thought to be due to the fracture of a highly heterogeneous system that surrounds
 the family of  asperities. A catastrophic EQ of magnitude Mw = 6.3 occurred on 06/04/2009 in central Italy. The majority 
of the damage occurred in the city of L'Aquila. Clear kHz - MHz EM anomalies have been detected prior to the L' Aquila EQ. Herein, we investigate the seismogenic origin of the detected MHz anomaly. The analysis in terms of intermittent dynamics of critical fluctuations reveals that the candidate EM precursor: (i) can be described in analogy with a thermal continuous phase transition; (ii) has anti-persistent behaviour. These features suggest that the emerged candidate precursor could be triggered by microfractures in the highly disordered system that surrounded the backbone of asperities of the activated fault. We introduce a criterion for an underlying strong critical behavior. In this field of research the reproducibility of results is desirable: the results should be verify by a number of precursory MHz EM emissions. We refer to precursory MHz
 EM activities associated with nine significant EQs that occurred in Greece. We conclude that all the MHz EM precursors under study also can be described in analogy with a continuous second-order phase transition showing antipersistent behaviour. 

\end{abstract}

\maketitle

\section{Introduction}

EQs are large-scale fracture phenomena in the Earth's heterogeneous crust. A vital problem in material
science and geophysics is the identification of precursors of macroscopic defects or shocks. In physics, the 
degree to which we can predict a phenomenon is often measured by how well  we understand it. However, despite  the large amount of experimental data and the considerable effort that has been undertaken by the material scientists, many questions about fracture processes remain standing. Especially, many aspects of EQ generation still escape our  full understudying. This fact is reflected in the disappointing progress on short-term EQ prediction [1]. 

Fracture induced physical fields allow a real-time monitoring of damage evolution in materials during mechanical loading. EM emissions in a wide frequency spectrum ranging from kHz to MHz are produced by opening cracks, which can be considered as the
so-called precursors of general fracture. These precursors are detectable both at a laboratory and a geophysical scale [1-12]. Our main observational tool is the monitoring of the fracture which occurs in the focal area before the final break-up by recording their kHz-MHz EM emissions.

Since 1994, a station was installed at a mountainous site of Zante island $(37.76^{o}N-20.76^{o}E)$ in the Ionian Sea (western Greece) with the following configuration: (i) six loop antennas detecting the three components (EW, NS, and vertical) of the variations of the magnetic field at 3 kHz and 10 kHz respectively; (ii) two vertical $\lambda/2$ electric dipoles detecting the electric field variations at 41 and 54 MHz respectively and (iii) two Short Thin Wire Antennas (STWA) of 100 m length each, lying on the Earth's surface, detecting ultra low frequency (ULF) ($<~ 1~ Hz$) anomalies, at EW and NS direction respectively. The 3 kHz, 10 kHz, 41 MHz, and 54 MHz were selected in order to minimize the effects of the sources of man-made noise in the mountain area of the Zante island. All the EM time-series were sampled at 1 Hz. Such an experimental setup helps to specify not only whether or not a single EM anomaly is preseismic in itself, but also whether a combination of such disturbances at different frequencies can be characterized as pre-seismic. 
 
Clear EM anomalies have been detected over periods ranging from a few days to a few hours prior to recent destructive EQs in Greece, and it is by now recognized that the pre-fracture kHz-MHz EM time series contain valuable information about the fracture EQ preparation process [2-11]. Being non-destructive, monitoring techniques based on  these fracture-induced fields are a basis for a fundamental understanding of fracture mechanism and for developing consecutive models of rock/focal area behaviour [12]. An improved understanding of the kHz-MHz EM precursors, especially its physical basis, has direct implications for the study of EQ generation processes and EQ prediction research. There is a growing number of indications on possible coupling between ionospheric anomalies and EQs [3]. 
The appearance of the ULF anomaly verifies that a  physically powerful fracto-electrification, and thus the fracture process, has been extended up to the surface layer of the crust. 

We emphasize that  the MHz radiation appears earlier than the kHz activity [1, 10]. This situation is 
in harmony with laboratory results [1]. An important challenge in this field of research is to distinguish characteristic epochs in the evolution of precursory EM activity and identify them with the equivalent last stages in the EQ preparation process. Recently, we have proposed the  following two epochs/stages model of EQ generation [5]:

(i) The abrupt emergence of strong impulsive kHz EM emission in the tail of the precursory EM activity is triggered by the fracture of the backbone 
of high strength and large asperities  that are distributed along the activated fault sustaining the system. The analysis
 of the detected precursory kHz EM activities indicate an underlying non-equilibrium process without any footprint of an equilibrium second-order phase transition. Notice, these precursory time series show persistent behavior, i.e., if the amplitude of EM fluctuations increases in a time interval, it is likely to continue increasing in the interval immediately following. The persistency implies a nonlinear positive feedback in the fracto-EM mechanism. The existence of a positive feedback justifies
 the absent of any signature of a second order transition. Based on well documented scaling properties of fault surface topology, we argue that the candidate kHz EM precursors might be originated during the slipping of two rough and rigid fractional-Brownian-motion-type profiles one over the other, with a roughness which is consistent with field and laboratory studies [7]. The results also suggest that the activation of a single EQ (fault) is a reduced self-affine image of the whole regional  seismicity and a magnified self-affine image of the laboratory seismicity [7].
 The last suggestion is also supported by arguments based on non-extensive (Tsallis) statistical mechanics [8,9].

(ii) The initial MHz anomaly is thought to be due to the fracture of a highly heterogeneous system that
 surrounds the family of  asperities [7]. The associated arguments have been supported by means of criticality, 
in particular in terms of  the recently introduced method of critical fluctuations and Levy statistics [11]. 
More precisely,  the analysis indicates that the MHz EM activity is well described in analogy with a thermal
 continuous phase transition [5]. A fractal spectral analysis by means of
 Hurst exponent shows that this activity 
has antipersistent properties, i.e., if the EM  fluctuations increase in period, they are likely to continue
 decreasing in the interval immediately following and vice versa [5, 10]. An antipersistent behavior implies a set 
of fluctuations  tending to induce a stability within the system, i.e., the existence of a
 nonlinear negative 
feedback in the underlying
 fracto-EM mechanism which ``kicks'' the opening cracks away from extremes. The observed antipersistency, 
which is
 rooted in the heterogeneity of the system [5, 10], enables the fracture in highly heterogeneous systems
 to be described via an analogy with thermal continuous phase transitions.  We have focused especially  
on the naturally arising question: what is the physical mechanism that organizes the heterogeneous
 system in its critical state? Combining ideas of Levy statistics, Tsallis statistics, and criticality 
on one hand and features hidden in the precursory MHz time series on the other hand, we argue that a Levy walk
 type mechanism can organize the heterogeneous system to criticality [11]. Moreover, based on a numerically
 produced truncated Levy walk, we propose [11] a way to estimate in the 
stage of critical fluctuations: (i) the associated Levy index-a, which describes quantitatively the underlying Levy
 dynamics, and (ii) the range of values where the nonextesitive Tsallis index q is restricted [11].

A catastrophic EQ of magnitude Mw = 6.3 occurred on April 6, 2009 (01h 32m 41s UTC) in central Italy. The majority of the damage occurred in the city of L'Aquila. Notice, this event was very shallow, namely, its depth was a few kilometres. Clear EM anomalies from ULF, kHz up to MHz have been detected prior to the L' Aquila EQ. Herein, we focus on the MHz EM anomaly that emerged on March 26 1999. The main goal of this contribution is to investigate the possible seismogenic origin of the detected anomaly in terms  of the above mentioned two stages / epochs model. More precisely, we attempt to verify that the detected MHz EM activity: (i) is well described in analogy with a thermal continuous phase transition, and (ii) is  characterized by Hurst exponents that indicate the existence
 of an underlying antipersistent fracture mechanism. The possible appearance of these two features taken together support the hypothesis that the recorded precursor 
coud be rooted in the fracture of the heterogeneous medium that surrounded the family of strong asperities that were distributed along the activated fault of the L'Aquila EQ. The analysis reveals that the aforementioned two expected features are hidden in the precursor under study.   

The physics of EQs has been demonstrated to be a very complicated matter, and the road seems to be 
long and hard. A difficulty  stems from the lack of large and reliable database: most studies in this  area have been 
limited by a lack of enough experimental results to conduct a statistically significant analysis of the phenomena and
obtain firm results. In this field of research the reproducibility of results is really desirable: the results should be verified by a numerous of precursory EM emissions. However, the collection of such a volume of appropriate EM data for statistical purposes requires some decades of years at least. Indeed, due to their absorption the EM precursors are associated with surface EQs of magnitude approximately 6 or greater that occur on land or closed to coastline . In the present study we investigate candidate precursory MHz EM activities associated with other 9 EQs of magnitude approximately 6 or greater that occur on land or closed to coastline in Greece. The results are in surprising agreement with those found in the study of the precursor associated with the L'Aquila EQ. Indeed, all the MHz EM precursors can be described in terms of a continuous second order phase transition and show antipersistent behavior. 

We emphasize that kHz EM anomalies have also recorded prior to the L'Aquila EQ on April 4, 2009. The performed multidisciplinary analysis supports the consideration 
that this precursory EM activity could be triggered by the fracture of a backbone of high strength and large asperities 
that are distributed along an activated fault. Finally, the appearance of a clear ULF anomaly during the last days prior to the L'Aquila EQ strongly verifies that a  physically powerful fracto-electrification, and thus a fracture process, has been extended up to the surface layer of the crust. We recall that the L'Aquila EQ was very shallow. The successive launch of ULF, kHz and MHz EM anomalies during a few days prior to the L'Aquila EQ  helps to specify not only whether or not a single EM anomaly is preseismic in itself, but also whether a combination of such disturbances at different frequencies could be characterized as pre-seismic. A relevant manuscript is under preparation.

The organization of this work is the following :  In section 2  the method of critical fluctuations (MCF) [13, 14] is  presented. In section 3 we apply 
the MCF to the detected  MHz EM animaly prior to the L'Aquila EQ. In section 4 the reproducibility of results is investigated. In section 5 the connection between the antipersistency and criticality is demonstrated. Finally, in section 6 we discuss our conclusions.

\section{The method of critical fluctuations}

The MCF, which constitutes a statistical method of analysis for the critical fluctuations in systems that undergo a continuous phase transition at equilibrium, has been introduced, recently [13, 14]. The authors have shown that the fluctuations  $\phi $ of the order parameter that corresponds to successive configurations of critical systems at 
equilibrium obey a dynamical law of  intermittency, which can be described in terms of a 1-D nonlinear map. The invariant density $\rho(\phi)$ for such a map is characterized by a plateau which decays in a super-exponential  way (see Fig. 1 in [13]). Importantly, the exact dynamics at the critical point can be determined analytically for  a large class of critical systems introducing the so-called critical map [13]. For small values of $\phi $, this critical map can be approximated as  

\begin{equation}
\phi_{n+1}=\phi_n + u \phi_n^z +\epsilon_n .
\label{eq:eq1}
\end{equation}

The shift parameter $\epsilon_n$ introduces a non-universal stochastic noise: each physical system has its characteristic ``noise'', which is expressed through the shift parameter $\epsilon_n$.  Notice, for thermal systems the exponent z is connected with the isothermal critical  exponent $\delta$ as $z= \delta + 1$. We emphasize that the plateau region of the invariant density $\rho(\phi)$ corresponds to the laminar region of  the critical map, where fully correlated dynamics takes place. The laminar region ends when the second nonlinear term in Eq. (1) becomes relevant. However, due to the fact that the dynamical law (1) changes continuously with $\phi $, the end of the laminar region cannot be easily defined based on a strictly quantitative criterion. Thus, the end of the laminar region should be generally treated as a varying parameter.

Based on the aforementioned description of the critical fluctuations, the MCF develops an algorithm permitting the extraction of the critical fluctuations, if any, in a measured time series. The crucial observation in this approach is the fact that the distribution $ P(l) $ of the laminar lengths $l$ of the above
 mentioned intermittent map (1) in the limit  $\epsilon_n \to  0 $ is given by the power law [14]

\begin{equation}
P(l) \sim l^{-p_l},
\label{eq:eq2}
\end{equation}

where the exponent $p_l$ is connected with the exponent $z$ via $p_l=\frac{z}{z-1}$. 

Therefore the exponent $p_l$ is connected with the isothermal exponent $\delta$ as follows:

\begin{equation}
p_l= 1 + \frac{1}{\delta}
\label{eq:eq3}
\end{equation}

Inversely, the existence of a power law such as in Eq. (2), accompanied by a plateau form of the corresponding density $\rho(\phi)$, is a signature of underlying correlated dynamics similar to the critical behavior [13, 14].
  
We give emphasis to the fact that it is possible in the frame of universality, which is characteristic for continuous second-order transitions, to give meaning to the exponent $p_l$ beyond the thermal phase transitions [14].

The MCF is directly applied to time series  or to segments of time series which appear a cumulative type stationary behavior. The main aim of the MCF is to estimate the exponent $p_l$. The distribution of the laminar lengths of fluctuations included in a stationary window is fitted by the function:

\begin{equation}
P(l) \sim  l^{-p_2} e^{-p_3 l} 
\label{eq:eq4}
\end{equation}

We focus on the exponents $p_2$ and $p_3$.  If the exponent $p_3$ is zero, then, the exponent $p_2$ is equal  to the exponent $p_l$.  Practically, as the exponent $p_3$ approaches to zero, then, the exponent $p_2$ approaches to $p_l$,  while, the laminar  lengths tend to follow a power-law type  distribution.  We note that Eq. (3) suggests that the  exponent $p_l$ ( or $p_2$) should be greater than 1. In conclusion, the critical profile of the temporal fluctuations is restored in the restrictions:  $p_2 > 1$ and $p_3 \approx 0$.

We stress that when the exponent $p_2$ is smaller than one, then, independently of the $p_3$-value, the system is not  at a critical state. Generally, the exponents $p_2, p_3$ have a competitive character, namely, when the exponent $p_2$ decreases the associated exponent $p_3$ increases (they are mirror image each to other).
In this way, 
we can identify the deviation  from the critical state. To be more concrete, as the exponent $p_2$ ($p_2<1$) is close to 1 and simultaneously the exponent $p_3$ is close to zero,  then, the system is at a sub-critical state. As the system removes from the critical state, the exponent $p_2$ decreases while simultaneously the exponent $p_3$ increases reinforcing, in this way, the exponential character of the laminar  lengths distribution. A situation like this indicates that the dynamics of the system becomes rather chaotic than deterministic [15, 16]. 

In summary, the research of criticality in natural systems can be quantitatively accomplished by estimating the values of only two parameters, namely the exponents $p_2$ and $p_3$. 

Up to now, the MCF has been applied on numerical experiments of thermal systems (Ising models) [14], on electromagnetic pre-seismic signals [5], and on electrocardiac signals from biological tissues [17].

Comment: The end of laminar regions is a variable parameter inside the decay region. In a recent work [18], it has been presented a prototype intermittent map of the form (1) which was in agreement with the provided properties. In the frame of this prototype, the exponent $p_2$ had almost constant value for all
the end points of laminar regions. This means that every trajectory carry the correct information of the nonlinear term. This property could be characterized as a completeness behavior which notices a stable character of criticality. 
This behavior is rarely observed in real time series and so, we consider as $p_2$ exponent the maximum value because for this $p_2$-value the exponent $p_3$ has its minimum value and so the $p_2$ is closest to the exponent of criticality $p_l$.

Based on the above mentioned considerations, the case where the  $p_2$-values  are greater  than 1 and the corresponding $p_3$-values close to zero for all end points of laminar regions indicates that all the trajectories are critical and carry almost the same information about the dynamical term. We can say that in this case the system is characterized by a "strong criticality".

\section{Application of the MCF method to the detected pre-seismic MHz EM radiation prior to the L'Aquila earthquake}

As it was said, a catastrophic EQ of magnitude Mw = 6.3 occurred on April 6, 2009 (01h 32m 41s UTC) in central Italy. 
The majority of the damage occurred in the city of L'Aquila. Notice, this event was very shallow, namely, its depth 
was a few kilometres. Clear EM anomalies from ULF, kHz up to MHz have been detected prior to the L' Aquila EQ.
 Herein, we focus on the 41-MHz EM anomaly that emerged prior to the L'Aquila EQ, i.e., on March 26 2009.
 The main goal of this contribution is to investigate the possible seismogenic origin of the detected anomaly in terms of the above mentioned two stages / epochs model [5]. The data were sampled at 1 Hz. 

It is significant to isolate the window of the time series carrying the footprints of the critical state. Our analysis verified the existense of a critical window (CW). Indeed, we have identified a characteristic time interval including 15000 points (Fig. 1a) in which a cumulative stationary condition occurs. As mentioned above, the existense of cumulative stationary behaviour in the candidate CW is a necessary condition for applying the MCF method to a nonequilibrium process. The stationary behaviour of the CW has been checked by estimating the mean value and the standard deviation for various time intervals in the CW, all having a common origin. The results are shown in Fig. (1b). The corresponding distribution of the amplitude $P(\psi)$ of the emerged EM pulses in the candidate CW is shown in Fig.2. It is characteristic the appearence of the plateau region in the top of distribution, as it is provided for the invariant
density of critical map [13]. The next step is to produce the distribution $ P(l) $ where the length l is the laminar interval, i.e., the stay time
 within the laminar region. More precisely, the laminar lengths are  described by the lengths of the sub-sequences 
in the EM time series that are resulted  from successive $\psi$-values  obeying  the condition $\psi_o \leq \psi  \leq  \psi_l$,  where $\psi_o $ is the fixed point
 and the end of the laminar region $\psi_l$ is a variable parameter. The more abrupt decay side of the distribution $P(\psi)$ is usally the fixed point $\psi_o$ [15]  but in the case of a symmetric distribution, like here, we select as fixed point the beggining of the distribution  (here $ \psi_o = 650 $). Using the fitting function  (4)
we estimate the exponents $p_2, p_3$ for different $\psi_l$- values. In Fig. 3a the exponents $p_2, p_3$ vs the end point $\psi_l$ are shown. In order to avoid end effects we ignore  the exponents values coming from the end points of the distribution $P(\psi)$.  As it results from Fig. 3, this
 window can be characterized as a strong CW. Indeed, the majority of the $p_2$-values are restricted in a zone from 1.20 to 1.43, while the exponent $p_3$ has values close to zero. Notice, that the quality of fitting in fig.3b  was excellent ($R^2 > 0.99$). 

In fig.4, for comparison reasons, we present a typical distribution $ P(l) $ for a time interval (28/3/2009) regarded as noise which does not exhibit  criticality. 

In summary, based on the above mentioned results in terms of MCF, we argue that the candidate critical window under study in the preseismic time series really is a critical window reflecting an underlying fracture process which is characterized by "strong criticality". The underlying dynamics posses intermittent characteristics, shown to describe the temporal fluctuations of the order parameter of a macroscopic system at its critical point.

\begin{figure}\includegraphics*[width=8 cm]{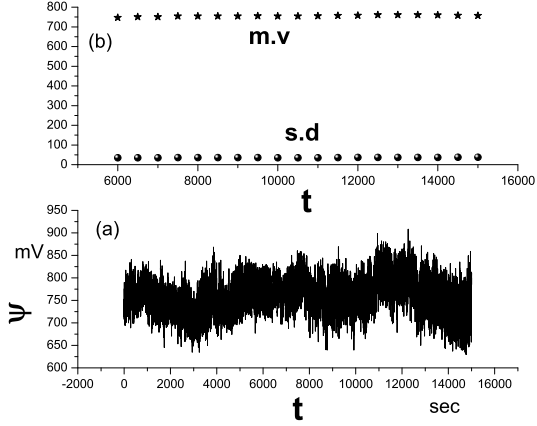}
\caption{\label{fig:fig1} The case of the L'Aquila EQ. (a) The MHz EM precursory fluctuations window .
 This window has a duration of 15000 points (1p/sec) and appeared almost 3.5 days before the EQ occurence.
(b) The temporal evolution of the cumulative mean value  and  standard deviation  for increasing number of points 
starting from an initial set of 6000 points.}
\end{figure}

\begin{figure}
\includegraphics*[width=8 cm]{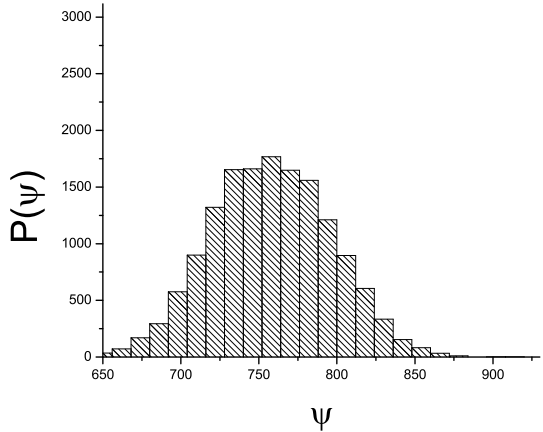}

\caption{\label{fig:fig2} The distribution $P(\psi)$ for the values of timeseries 1(a)  is shown .}
\end{figure}

\begin{figure}
\includegraphics*[width=8 cm]{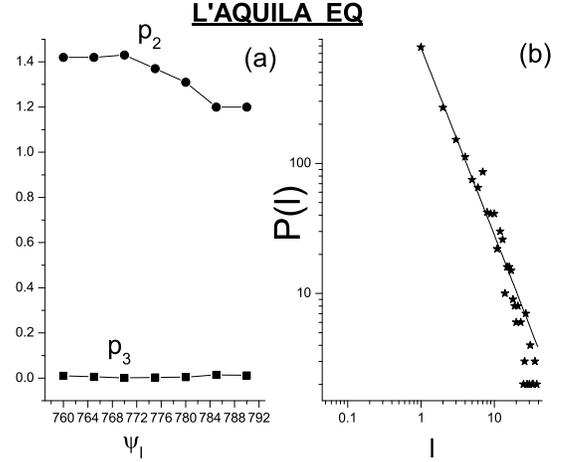}

\caption{\label{fig:fig3} (a)  Case of the L'Aquila EQ. The exponents $p_2,p_3$ vs $\psi_l$ indicate an underlying
 strong critical behavior.
 The corresponding critical window has a length of 15000 points (1p/sec) and is appeared almost 3.5 days before
 the EQ occurence. The $p_2$-values are restricted in the zone  (1.20--1.43).
 (b) The distribution of the laminar lengths, as it is produced by the analysis in terms of MCF,
for end point $\psi_l=770$. The fitting function 
gives $p_2=1.43$ and $p_3=0.0008$.
For concrete end point the distribution $P(l)$ is
more close to a power-law ($p_3$ minimum and $p_2$ maximum ). 
The solid line represents the fit of data by the fitting function (Eq.~4). The quality of fitting 
is very nice ($R^2=0.996$).}

\end{figure}

\begin{figure}

\includegraphics*[width=8 cm]{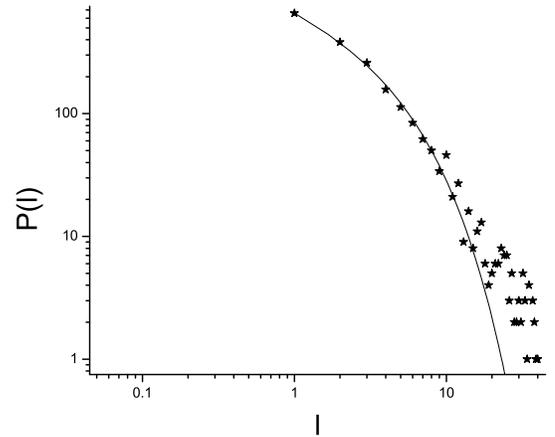}

\caption{\label{fig:fig4} The distribution of the laminar lengths, as it is produced by the analysis in terms of MCF, 
from a segment of preseismic EM signal of L'Aquila (28/3/2009) which does not exhibit criticality. The fitting
function gives $p_2=0.48$ and $p_3=0.22$. }

\end{figure}

\section{The reproducibility of results}

In this field of research the reproducibility of results is desirable: 
the results should be verified by a number of precursory EM emissions. Thus, a question effortlessly arises is whether the appearance of windows showing strong criticality, as it has been described previously, is systematically observed in candidate precursory MHz EM activities associated with significant EQs that occurred  in Greece during the last years. In the present section we refer to MHz EM activities recorded prior to nine significant EQs that occurred in land or near the coast line of Greece. We note that it is possible to have
the existence of more than one critical window in the preseismic EM fluctuations. In the following, for statistical reasons, we present the  critical window which has the greater duration.

(a) The Athens EQ. On 7 September 1999 (11:56:49 UT) the Athens EQ (38.12$^\circ$N, 23.60$^\circ$E) occurred with magnitude $M = 5.9$ [5, 7,10]. In figs. 5a,6a the associated CW has a duration 5.5h and it is appeared almost 4 days before the EQ event.

(b) The Kozani-Grevena EQ. On 13 May 1995 (8:47:13 UT) the Kozani-Grevena EQ  (40.17$^\circ$N, 21.68$^\circ$E) occurred with magnitude $M = 6.6$. 
In figs. 5b,6b the associated CW has a duration 8h and it is appeared almost 2 days before the EQ event.

(c) The Zante EQ. On 11 April 2006 (17:29:28 UT) the Zante EQ (37.68$^\circ$N, 20.91$^\circ$E) occurred 
close to the Zante island with magnitude 5.9. In figs. 5c,6c the associated CW has a duration 5h and it 
is appeared almost 19h before the EQ event.

(d) The Methoni EQ. On 14 February 2008 (10:09:23 UT) the Methoni 
EQ (36.55$^\circ$N, 21.77$^\circ$E) with a magnitude $M=6.9$ occurred. In figs. 5d,6d the associated CW has 
a duration 1.5d and it is appeared almost 6 days before the EQ event.

(e) The Andravida EQ. On 8 June 2008 (12:25:28 UT) the Andravida EQ (37.95$^\circ$N, 21.53$^\circ$E) with a magnitude $M=6.4$ occurred. In figs. 5e,6e the associated CW has a duration 5h and it is appeared almost 2 days before the EQ event.

(f) The Lamia EQ. On 13 December 2008 (11:27:19 UT)  the Lamia EQ (38.72$^\circ$N, 22.5$^\circ$E) with a magnitude $M=5.7$ occurred. In figs. 5f,6f the associated CW has a duration 6h and it is appeared almost 20h before the EQ event.

(g) The Kefalonia EQ. On 25 March 2007 (13:57:58 UT)  the Kefalonia  EQ (38.34$^\circ$N, 20.42$^\circ$E) with a magnitude $M=6$ occurred. In figs. 5g,6g the associated CW has a duration 5h and it is appeared almost 3 days before the EQ event.

(h) The Agios Nikolaos EQ. On 13 Junuary  2009 (06:12:43 UT)  the  Agios Nikolaos  EQ (35.66$^\circ$N, 26.39$^\circ$E)  with a magnitude
$M=5.7$  occurred. In figs. 5h,6h the associated CW has a duration 8h and it is appeared almost 5 days before the EQ event.

(i) The Strofades EQ. On 16 February  2009 (23:16:38 UT)  the  Strofades  EQ (37.13$^\circ$N, 20.78$^\circ$E)  with a magnitude
$M=6$  occurred. In figs. 5i,6i the associated CW has a duration 5.5h and it is appeared almost 10 days before the EQ event.

We clarify that the precursors associated with the above mentioned (a)-(f) and (i) seismic events have been recorded by the sensors of the Zante station, while those associated with the (g) and (h) shocks have been detected by the sensors of the Neapolis (east Crete) station (35.26$^\circ$N, 25.61$^\circ$E)  and Kefalonia station ((38.18$^\circ$N, 20.59$^\circ$E), respectivelly. 

Figs. 5 and 6 importantly show that in all the cases the EM fluctuations included within the critical windows are fitted by the function  (4)  
with exponent $p_2$ having values greater than 1 and exponent $p_3$ having values close to zero, while the quality of fittings were excelent ($R^2 > 0.99$). Based on this result, we conclude that in the EM precursor associated with the L'Aquila EQ are hidden features of "strong criticality" similar to those hidden in EM
 precursors associated with other nine seismic events. Thus the reproducibility of results is satisfied in 
ten different MHz EM precursors. 

\begin{figure}

\includegraphics*[width=8 cm]{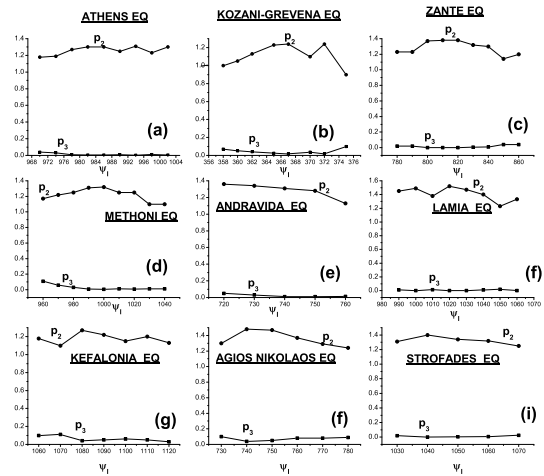}

\caption{\label{fig:fig5} The exponents $p_2,p_3$ vs $\psi_l$ indicate an underlying strong critical behavior
for the other 9 EQ events  as these are described in text. The corresponding CW are the greater in duration for each EQ. }
\end{figure}

\begin{figure}
\includegraphics*[width=8 cm]{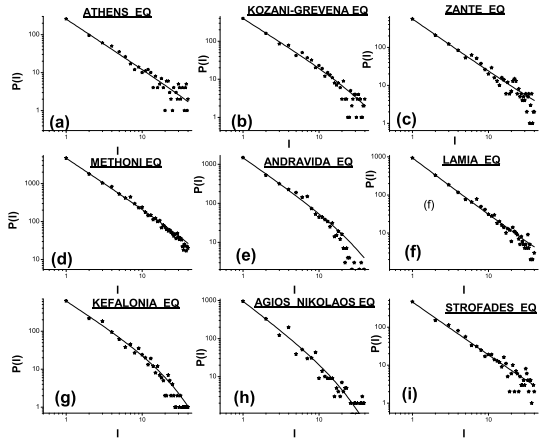}

\caption{\label{fig:fig6} The distribution of the laminar lengths, as it is produced by the analysis
 in terms of MCF, from the other 9 EQ events assosiated to the corresponding critical windows of Fig.5. 
All the distributions are more close
to power-law, for each case. }

\end{figure}

\section{The persistent behaviour of the candidate electromagnetic precursor}

The Hurst exponent $H$ characterizes the persistent/anti-persistent properties of the signal [19-20]. The range $0<H<0.5$ indicates an anti-persistency reflecting that increases in the value of a time series are likely to be followed by decreases and conversely. Physically, this implies a set of fluctuations tending to induce a stability to the system, namely a non-linear feedback mechanism that ``kicks'' the opening rate away from extremes. On the contrary, the range $0.5<H<1$ indicates persistent behavior. This means that increases in the value of a time series are likely to be followed by increase, namely the system has been starting to self-organize by a positive feedback process.

In Ref. [5]  we have introduced the following connection between the exponents $H$ and $p_2$:

\begin{equation}
H=2-\frac{3}{2} p_2 .
\label{eq:eq5}
\end{equation}

Eq. (5) suggests that the allowed range of 
$p_2$ values for a second-order phase transition, i.e., $p_2 >1 $, leads to the 
condition $ H<0.5 $ which indicates an underlying antipersistent mechanism.  On the other hand, Eq. (5), as it has been
found [5] is valid only if the exponent $p_2$ lies in the interval $1<p_2<1.5$. The present analysis systematically leads to exponents $p_2$  which are in  agreement with the above limits of $p_2$ for all the 10 EQs under study. 

To summarize, in the EM precursor possibly associated with the L'Aquila EQ are hidden features of strong criticality and antipersistency similar to those hidden in the EM precursors associated with other 9 seismic events. Notice, the interplay between the heterogeneities in the pre-focal area and the stress field could be responsible for the observed antipersistent pattern [5, 10 ].

\section{Conclusion-discussion}

Fracture induced EM fields ranging from kHz to MHz allow a real-time monitoring of damage evolution in materials during  mechanical loading. These EM precursors are detectable both at a laboratory and a geological scale containing valuable information about the fracture / EQ preparation process. MHz EM anomalies have been recorded prior to the catastrophic L'Aquila EQ a few days before the EQ occurrence. EM anomalies possibly associated with the preparation of EQs would be recognized as real precursors only  when the physical mechanism of their origin would be clarified. Thus, an important challenge in this field of research is to connect the detected precursory EM activity with a last stage in the EQ preparation process. Previous studies have suggested that the systematically detected MHz EM activity prior to significant EQs could be originated during cracking in the highly heterogeneous material that surrounds the backbone of large and strong entities distributed along the activated fault sustaining the system. We find that the candidate EM precursor associated with the L'Aquila EQ is in harmony with this suggestion. Indeed, in analogy to the study of critical phase transitions in statistical physics, it has been proposed that the fracture of heterogeneous materials could be viewed as a critical phenomenon [21-22]. Based on a recently presented statistical method of analysis for the critical fluctuations in systems which undergo a continuous phase transition at equilibrium [13-14],  we show [5] that the detected MHz anomaly prior to the L'Aquila EQ can be described in analogy with a thermal continuous phase transition by means of a recently introduced method of critical fluctuations. Especially, we introduce a criterion for an underlying strong critical behavior. According to this criterion the majority of trajectories in the properly defined laminar region, ``carries'' the information of an underlying criticality. The precursor under study follows this criterion. Importantly, the precursor shows antipersistent behaviour indicating an underlying non-linear feedback system that ``kicks'' the opening rate of cracks away from extremes. The antipersistent behavior is in agreement with a system which undergo a continuous phase transition at equilibrium.  In this field of research the reproducibility
 of results is desirable. We study precursory MHz EM activities associated with nine significant EQs that occurred 
in Greece during the last years. We show that all these nine MHz EM precursors behave as that detected prior to 
the L'Aquila EQ: they can be described in analogy with a continuous second-order phase transition showing strong
 criticality and antipersistent behaviour. The thesis underlying the present effort is that the emerged MHz EM anomaly
 prior the L'Aquila EQ  could be emitted by the fracture of disordered material in the focal area 
of this EQ that 
surrounded the family of asperities of the activated fault.

\end{document}